\documentclass[journal,onecolumn]{IEEEtran}
\pdfoutput=1 
\usepackage{graphicx} 
\usepackage{amsmath} 
\usepackage{amssymb}
\usepackage{cite}
\usepackage{amsfonts}
\usepackage{caption}
\usepackage{subcaption}
\usepackage{psfrag}
\usepackage{array}
\usepackage{multirow}
\usepackage{multicol}
\usepackage{mathtools, cuted}
\usepackage{setspace}
\usepackage{longtable}
\usepackage{float}
\usepackage{color}
\usepackage{algorithm}
\usepackage{algorithmic}
\usepackage[section]{placeins}

\newcommand{\bm}{\mathbf}
\newcommand{\bsy}{\boldsymbol}
\newcommand{\be}{\begin{equation}}
\newcommand{\ee}{\end{equation}}
\newcommand{\bea}{\begin{eqnarray}}
\newcommand{\eea}{\end{eqnarray}}

\newcommand{\br}{{\bm r}}

\newcommand{\bA}{{\bm A}}
\newcommand{\bI}{{\bm I}}
\newcommand{\bW}{{\bm W}}
\newcommand{\bE}{{\bf E}}
\newcommand{\bF}{{\bf F}}
\newcommand{\bD}{{\bf D}}
\newcommand{\bC}{{\bf C}}
\newcommand{\bB}{{\bf B}}
\newcommand{\bS}{{\bf S}}
\newcommand{\bH}{{\bf H}}

\newcommand{\bR}{{\bf R}}

\newcommand{\bV}{{\bf V}}
\newcommand{\bU}{{\bf U}}

\newcommand{\bd}{{\bf d}}

\newcommand{\bs}{{\bf s}}
\newcommand{\bx}{{\bf x}}

\newcommand{\n}{{\bm n}}

\newcommand{\bGamma}{\mbox{\boldmath$\Gamma$}}

\newcommand{\snri}{\mbox{$\frac{\sigma_n^2}{\sigma_d^2}$}}
\newcommand{\complex}{\mathbb{C}}
\renewcommand{\natural}{\mathbb{N}}

\newcommand{\channelequalizationmatrix}{\bH\bH^{\rm \dagger}+\snri \bI}
\newcommand{\channelequalizationmatrixop}{\mathbf{\Psi}}
\newcommand{\HH}{\bH\bH^{\rm \dagger}}
\newcommand{\delaymatrix}{\bm{\Pi}}
\newcommand{\dopplermatrix}{\bm{\Delta}}
\newcommand{\delayindex}{l}
\newcommand{\dopplerindex}{k}
\newcommand{\bT}{\mathbf{T}}
\newcommand{\bG}{\mathbf{G}}
\newcommand{\bL}{\mathbf{L}}
\newcommand{\bbL}{\mathcal{L}}
\newcommand{\bbU}{\mathcal{U}}
\newcommand{\bX}{\bm{X}}
\newcommand{\bY}{\bm{Y}}

\newcommand{\subcarrier}{\Delta f}
\newcommand{\kron}{\otimes}
\newcommand{\bn}{{\bf n}}
\DeclarePairedDelimiter{\ceil}{\lceil}{\rceil}
\begin{document}
	\title{Low complexity LMMSE receiver for OTFS}

	\author{\IEEEauthorblockN{Shashank Tiwari, Suvra Sekhar Das,~\IEEEmembership{Member,~IEEE} and Vivek Rangamgari \\}

		\thanks{Shashank Tiwari, Suvra Sekhar Das and Vivek Rangamgari are with G. S. Sanyal School of Telecommunications, Indian Institute of Technology, Kharagpur, India}
		\thanks{e-mail: shashankpbh@gmail.com, suvra@gssst.iitkgp.ernet.in and rkvivek97@gmail.com}}
		
		\maketitle
		
		\begin{abstract}
			Orthogonal time frequency space modulation is a  two dimensional (2D) delay-Doppler domain waveform. It uses inverse symplectic Fourier transform (ISFFT) to spread the signal in time-frequency domain. To extract diversity gain from 2D spreaded signal, advanced receivers are required. In this work, we investigate a low complexity linear minimum mean square error receiver which exploits sparsity and quasi-banded structure of matrices involved in the demodulation process which results in a log-linear order of complexity without any performance degradation of BER.
		\end{abstract}
	
		\IEEEpeerreviewmaketitle
	\section{Introduction}
	
	
	
Fifth generation new radio (5G-NR) \cite{3gpp38211} uses multi-numerology Orthogonal frequency division multiplexing (OFDM) system to cater to different requirements of 5G such as support for higher vehicular speed scenario and high phase noise. Although sub-carrier bandwidth in 5G-NR can be increased to combat Doppler spread, the provision of proportional decrement of CP length to retain OFDM symbol efficiency induces interference  when both delay and Doppler spreads are significant.  Orthogonal time frequency space modulation (OTFS) has been recently proposed in \cite{hadani_orthogonal_2017} to efficiently transfer data in such channel conditions. In OTFS, data symbols are spread across available time-frequency resources which can be exploited to extract diversity gain. Different receivers  have been proposed in the literature \cite{farhang_low_2018,murali_otfs_2018,raviteja_interference_2018,hadani_otfs:_2018,hadani_receiver-side_2019,li_low_2019}, which achieve such diversity gain.
	
	When exposed to a time variant channel (TVC), OTFS suffers from inter-symbol and inter-carrier interference \cite{raviteja_interference_2018}. Hence a simple matched filter receiver as in \cite{farhang_low_2018} is unable to suppress the interference sufficiently.  There can be two types of receivers, namely (i) linear receivers (LRx) and (ii) non-linear receivers (NLRx). NLRx (such as in \cite{murali_otfs_2018,raviteja_interference_2018,hadani_otfs:_2018,hadani_receiver-side_2019}) have near maximum likelihood (ML) performance but have iterative structure and high complexity. On the other hand LRx are simple in the structure but have relatively poorer performance than non-linear receivers. As linear processing requires inversion and multiplication of matrices, LRx still posses computational burden for OTFS as the time-frequency grid size in OTFS is very large.  Linear minimum mean square error (LMMSE) receiver which is well known for its interference cancellation capabilities \cite{hadani_otfs:_2018,jiang_performance_2011}, is extended to a low complexity form in this work.
	 
	
	Direct implementation of LMMSE receiver require complexity in the order of $O(M^3N^3)$, where $M$ and $N$ are total number frequency and time slots respectively. When the values of $M$ and $N$ are in order of 100's, the complexity of LMMSE receiver becomes extraordinarily large. To the best of the authors' knowledge, not much attention has been paid towards low complexity design of LMMSE receiver for OTFS in literature. We present a low complexity LMMSE receiver which has a complexity in the order of  $O(MN\log_2(N))$ without any degradation in BER performance.


 We use the following notations throughout the paper. We let $\bx$, $\bm X$ and $x$ represent vectors, matrices and scalars respectively. The superscripts $(-)^{\rm T}$and $(-)^{\rm \dagger}$ indicate transpose and conjugate transpose  operations, respectively. Notations $\bsy{0}$, $\bI_N$ and $\bW_L$ represent zero matrix, identity matrix with order $N$ and $L$-order normalized inverse discrete Fourier transform  (IDFT) matrix respectively. Kronecker product operator is given by $\kron$. 
 The operator $\mathrm{diag}\{\bm x\}$ creates a diagonal matrix with the  elements of vector $\bm x$.
Circulant matrix is represented by $circ\{\bm x\}$  whose first column is $\bm x$. Notations $E\{-\}$ and $\ceil{-}$ are expectation and ceil operators respectively. Column-wise vectorization of matrix $(\bm X)$ is represented by  $vec\{\bm X\}$. Natural numbers are denoted by $\natural$ . Complex conjugate value of $x$ is given by $\bar{x}$ whereas $j=\sqrt{-1}$.  
\section{System Model}
We consider an OTFS system with $M$ number of sub-carriers having $\subcarrier$ sub-carrier bandwidth and $N$ number of symbols having $T$ symbol duration. Total bandwidth $B=M\subcarrier$ and total duration $T_f=NT$. Moreover OTFS system is critically sampled i.e. $T\subcarrier=1$.

\subsection{Transmitter}
 QAM modulated data symbols, $d(k,l)\in \complex$, $k\in \natural[0~N-1]$, $l\in \natural[0~M-1]$, are arranged over Doppler-delay lattice $\Lambda=\{(\frac{k}{NT},~\frac{l}{M \subcarrier})\}$. We assume that $E[d(k,l) \bar{d}(k',l')] = \sigma_d^2 \delta(k-k',l-l') $, where $\delta$ is Dirac delta function. Doppler-delay domain data $d(k,l)$ is mapped to time-frequency domain data $X(n,m)$  on lattice $\Lambda^{\perp}=\{(nT,~m\subcarrier)\}$, $n\in \natural[0~N-1]$ and $m\in \natural [0~M-1]$ by using inverse symplectic fast Fourier transform (ISFFT). $X(n,m)$ can be given as \cite{hadani_orthogonal_2017},
\begin{equation}
X(n,m)=\frac{1}{\sqrt{NM}} \sum_{k=0}^{N-1}{\sum_{m=0}^{M-1}{d(k,l) e^{j2\pi [\frac{nk}{N}-\frac{ml}{M}]}}}.
\end{equation}
  Next, $X(n,m)$ is converted to a  time domain signal $s(t)$ through a Heisenberg transform as,
\begin{equation}
s(t)= \sum_{n=0}^{N-1}{\sum_{m=0}^{M-1}}{X(n,m)g(t-nT) e^{j2\pi m \subcarrier (t-nT)}},
\end{equation}
where, $g(t)$ is transmitter pulse of duration $T$. It has been shown in \cite{raviteja_practical_2019} that non rectangular pulse induces non-orthogonality which degrades BER performance. Thus, in this work, we assume a rectangular pulse i.e. $g(t)=\begin{cases}
1~ \text{if}~0\leq t \leq T \\
0 ~ \text{otherwise}
\end{cases}$.  

To obtain discrete time representation of OTFS transmission, $s(t)$ is sampled at the sampling interval of $\frac{T}{M}$ \cite{raviteja_practical_2019}. We collect samples of $s(t)$ in $\bs=[s(0)~s(1) \cdots s(MN-1)]$ and QAM symbols $d(k,l)$ are arranged in $M\times N$ matrix as,
\begin{equation}
\bD= \small{\begin{bmatrix}
	d(0,0) & d(1,0) & \cdots & d(N-1,0) \\
	d(0,1) & d(1,1) & \cdots & d(N-1,1) \\
	\vdots & \vdots & \ddots & \vdots  \\
	d(M-1,0) & d(M-1,1) & \cdots & d(N-1,M-1) \\
	\end{bmatrix}}.
\end{equation}
 Using above formulations, $\bs$ can be given as \cite{raviteja_practical_2019},
\begin{equation}
\bs = vec\{\bD \bW_N \}.
\end{equation}
Alternatively, if $\bd=vec\{\bD\}$, transmitted signal can also be written as matrix-vector multiplication,
\begin{equation}
\bs= \bA \bd,
\label{eqn:otfsmodmatrixvector}
\end{equation}
where $\bA= \bW_N \kron \bI_M $ is OTFS modulation matrix. Finally, a cyclic prefix (CP) of length $\alpha'\geq \alpha-1$ is appended at the starting of the $\bs$, where $\alpha$ is channel delay length described in Sec.~\ref{sec:systetmodel:channel}.
\subsection{Channel} \label{sec:systetmodel:channel}
We consider a time varying channel with $P$ paths having $h_p$ complex attenuation, $\tau_p$ delay and $\nu_p$ Doppler value for $p^{\rm th}$ path where $p\in \natural[1~P]$. Delay-Doppler channel spreading function can be given as,
\begin{equation}
h(\tau,\nu)=\sum_{p=1}^{P}{h_p \delta(\tau-\tau_p) \delta(\nu-\nu_p)}.
\end{equation}
The delay and Doppler values for $p^{\rm th}$ path is given as $\tau_p=\frac{\delayindex_p}{M\subcarrier}$ and $\nu_p=\frac{\dopplerindex_p}{NT}$, where $\delayindex_p\in \natural[0~M-1]$ and $\dopplerindex_p\in\natural[0~N-1]$ are dealy and Doppler bin number on Doppler-delay lattice $\Lambda$ for $p^{\rm th}$ path.  We assume that $N$ and $M$ are sufficiently large so that there is no effect of fractional delay and Doppler on the performance. We also assume the perfect knowledge of $(h_p,~\delayindex_p,~\dopplerindex_p)$, $p\in \natural[0~P-1]$, at the receiver as in \cite{farhang_low_2018,murali_otfs_2018,raviteja_interference_2018,hadani_otfs:_2018,hadani_receiver-side_2019,li_low_2019}. Let $\tau_{max}$ and $\nu_{max}$ be the maximum delay and Doppler spread. Channel delay length $\alpha= \ceil{\tau_{max}M\subcarrier}$ and channel Doppler length, $\beta= \ceil{\nu_{max} NT}$. Typically, $\alpha<< ~MN$ as well as $\beta<<~MN$ which dictates the system matrices to be sparse (as will be seen in Sec.~\ref{sec:structureofHH}). For example, if we consider an OTFS system with $\subcarrier ~= 15$ KHz, carrier frequency, $f_c=4$ GHz, $N=128$ and $M=512$. We take a 3GPP vehicular channel EVA \cite{series2009guidelines} with vehicular speed of 500 Kmph. Delay-Doppler channel lengths can be computed as, $\alpha=20<< ~MN~= 65536$ and $\beta=16<< ~MN~= 65536$.

\subsection{Receiver}
 After removal of CP at the receiver, received signal can be written as \cite{raviteja_practical_2019},
\begin{equation}
\br=\bH \bs +\bn,
\end{equation}
where, $\bn$ is white Gaussian noise vector of length $MN$ with elemental variance $\sigma_\n^2$ and $\bH$ is a  $MN\times MN$ channel matrix given as,
\begin{equation}
\bH=\sum_{p=1}^{P}{h_p \delaymatrix^{\delayindex_p} \dopplermatrix^{\dopplerindex_p}},
\label{eqn:channelmatrix}
\end{equation}
with $\delaymatrix=circ\{[0~1~0 \cdots 0]^{\rm T}_{MN\times 1}\}$ is a circulant delay matrix and $\dopplermatrix=\mathrm{diag}\{[1~ e^{j2\pi\frac{1}{MN}}~ \cdots e^{j2\pi\frac{MN-1}{MN}}]^{\rm T}\}$ is a diagonal Doppler matrix. We further process $\br$ through a LMMSE equalizer which results in estimated data vector,
\begin{equation}
\hat{\bd}=(\bH\bA)^{\rm \dagger} [(\bH \bA)(\bH \bA)^{\rm \dagger}+\snri \bI]^{-1} \br. 
\label{eqn:MMSEOTFS}
\end{equation}
\section{Low Complexity LMMSE Receiver for OTFS}
\begin{figure}[t]
	\centering
	\includegraphics[width=1\linewidth]{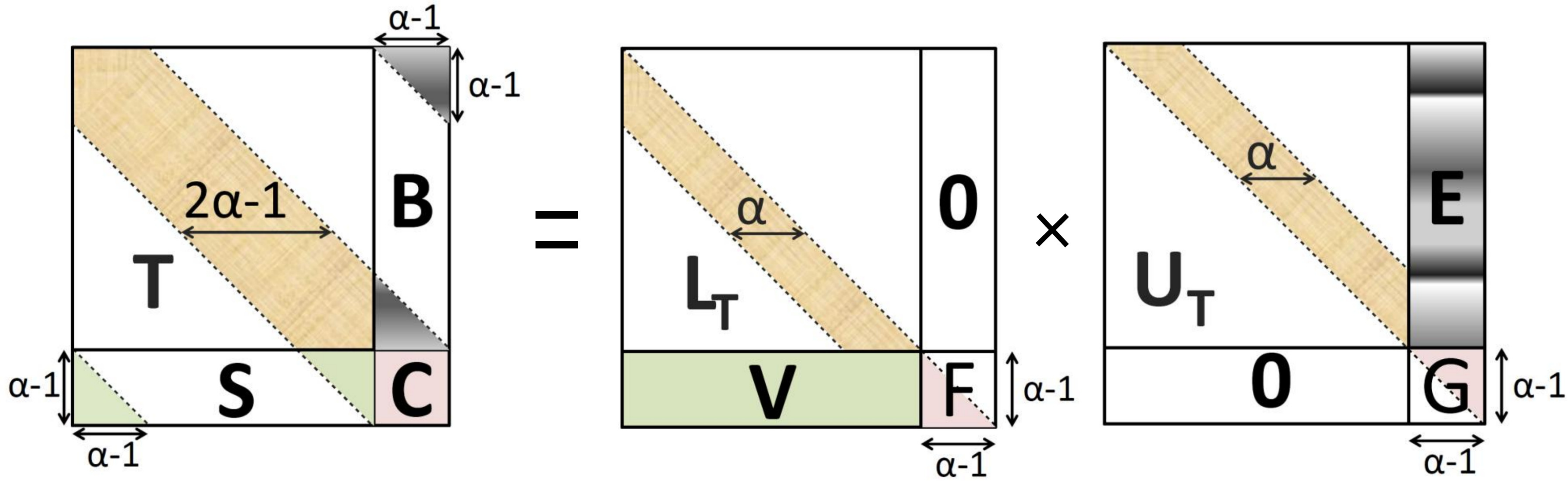}
	\caption{Structure of $\channelequalizationmatrixop=\bH \bH^{\rm \dagger}+\snri \bI$ matrix and its $LU$ factorization.}
	\label{fig:hh}
\end{figure}
When $g(t)$ is rectangular, $\bA$ in (\ref{eqn:otfsmodmatrixvector}) is unitary. Thus (\ref{eqn:MMSEOTFS}) is simplified to,
 \begin{equation}
 \hat{\bd}=\bA^{\rm \dagger}\underbrace{\overbrace{\bH^{\rm \dagger} [\bH\bH^{\rm \dagger}+\snri \bI]^{-1}}^{\bH_{eq}} \br}_{{\br_{ce}=\bH_{eq} \br}}.
 \label{eqn:MMSEOTFSsimplified}
 \end{equation}
 Thus LMMSE equalization can be performed as a two stage equalizer. In the first stage, LMMSE channel equalization is performed to obtain $\br_{ce}=\bH_{eq} \br$. Second stage is a OTFS matched filter receiver to obtain $\tilde{\bd}=\bA^{\rm \dagger} {{\br_{ce}}}$. We will show in Sec.~\ref{sec:lowcomplexity:computationofr} that the implementation of $\tilde{\bd}=\bA^{\rm \dagger} {{\br_{ce}}}$ is simple which requires $\frac{MN}{2}\log_2(N)$ complex multiplications (CMs). But direct implementation of $\br_{ce}=\bH_{eq} \br$ requires inversion of $\channelequalizationmatrixop=\channelequalizationmatrix$ and multiplication of $\bH^{\dagger}$ which need $O(M^3N^3)$ CMs. Thus, we need to reduce the complexity of $\br_{ce}=\bH_{eq} \br$. To do so, we investigate the structure of matrices involved in channel equalization in Sec.~\ref{sec:structureofHH}.
 \subsection{Structure of $\channelequalizationmatrixop =[\bH\bH^{\rm \dagger}+\snri \bI]$} \label{sec:structureofHH}
 
 Using (\ref{eqn:channelmatrix}), $\HH$ can be expressed as,
 \begin{equation}
 \HH= \sum_{p=1}^{P}{h_p \dopplermatrix^{\dopplerindex_p} \delaymatrix^{\delayindex_p}} \sum_{s=1}^{P}{\bar{h}_s \dopplermatrix^{-\dopplerindex_s} \delaymatrix^{-\delayindex_s}}.
 \end{equation}
 Since $\delaymatrix$ is a circulant matrix, it can be verified that $\delaymatrix^{\delayindex_p}= \bW \dopplermatrix^{-\delayindex_p} \bW^{\rm \dagger}$. Therefore,
 \begin{equation}
 \HH= \underset{p=s}{\sum_{p=1}^{P}}{|h_p|^2 \bI} + \underset{p\neq s}{\sum_{p=1}^{P}{\sum_{s=1}^{P}}}{h_p \bar{h}_s \delaymatrix^{\delayindex_p-\delayindex_s}\dopplermatrix^{\dopplerindex_p-\dopplerindex_s}}
 \label{eqn:closedformHH}
 \end{equation}
 Using (\ref{eqn:closedformHH}), $\channelequalizationmatrixop$ becomes,
 \begin{equation}
 \channelequalizationmatrixop={\underset{p=s}{\sum_{p=1}^{P}}{(|h_p|^2+\snri) \bI}} + {\underset{p\neq s}{\sum_{p=1}^{P}{\sum_{s=1}^{P}}}{h_p \bar{h}_s \delaymatrix^{\delayindex_p-\delayindex_s}\dopplermatrix^{\dopplerindex_p-\dopplerindex_s}}}.
 \label{eqn:closedformchest}
 \end{equation}
 
 Following (\ref{eqn:closedformchest}), it can be concluded that the maximum shift of diagonal elements in $\dopplermatrix$ can be $\pm (\alpha -1)$. Additionally due to the cyclic nature of shift, $\channelequalizationmatrixop$ is quasi-banded with bandwidth of $2\alpha-1$ as depicted in  Fig. ~\ref{fig:hh}. As discussed in  Sec.~\ref{sec:systetmodel:channel}, $\alpha << MN$, $\channelequalizationmatrixop$ is also sparse for typical wireless channel. 
Since, we need to implement $\channelequalizationmatrixop^{-1}$ in order to realize LMMSE receiver, we propose a low complexity LU decomposition of $\channelequalizationmatrixop$ in Sec.~\ref{sec:lowcomplexity}.
 \subsection{Low complexity LU factorization of $\channelequalizationmatrixop$} \label{sec:lowcomplexity}
 To implement the low complexity LU factorization of $\channelequalizationmatrixop$, we propose following partition of $\channelequalizationmatrixop$ ( by considering, $\theta=\alpha-1$ and $Q=MN-\theta$).

 \begin{equation}
 \underbrace{\begin{bmatrix}
\bT_{\footnotesize {Q\times Q}} & \bB_{\footnotesize { Q\times \theta}} \\
 \bS_{\footnotesize { \theta \times Q}} & \bC_{\footnotesize  {\theta \times \theta}}
 \end{bmatrix}}_{\channelequalizationmatrixop} = \underbrace{\begin{bmatrix}
 \bbL_{\footnotesize {Q\times Q}} & \bsy{0}_{\footnotesize { Q\times \theta}}  \\
 \bV_{\footnotesize { \theta \times Q}} & \bF_{\footnotesize { \theta\times \theta}}
 \end{bmatrix}}_{\bL} \times  \underbrace{\begin{bmatrix}
 \bbU_{{Q\times Q}} & \boldmath{\bE}_{\rm Q\times \theta}  \\
 \bsy{0}_{\rm \theta \times Q} & \bG_{\rm \theta\times \theta}
 \end{bmatrix}}_{\bU}
\label{eqn:partitionofHH}
 \end{equation}
Using the partition in (\ref{eqn:partitionofHH}), following equalities hold
\begin{eqnarray}
\label{eqn:T}
\bT&=&\bbL ~\bbU\\ \label{eqn:E}
\bE&=&\bbL^{-1}~\bB \\ \label{eqn:V}
\bV&=&\bS ~\bbU^{-1}\\  \label{eqn:FG}
\bF \bG &=& \bC-\bV \bE
\end{eqnarray}
Next we will discuss the solution of (\ref{eqn:T}-\ref{eqn:FG}) to compute $LU$ factorization of $\channelequalizationmatrixop$. Since $\bT$ is a banded matrix its LU decomposition can be computed using low complexity algorithm presented in \cite{walker_lu_1988}. $\bbL^{-1}~ \bB$ can be computed using forward substitution  algorithm for lower triangular banded matrix as explained in Algorithm~\ref{Alg:inverseoflowertriangularbandedmatrix}. We can compute (\ref{eqn:V}) in following two steps. As $\bbU^{\rm \dagger}$ is a lower triangular banded matrix, in the first step, we compute  $\bV^{\rm \dagger}= (\bbU^{\rm \dagger})^{-1} \bS^{\rm \dagger}$ using Algorithm~\ref{Alg:inverseoflowertriangularbandedmatrix}. Finally,  $\bV$ can be computed simply by taking hermitian of $\bV^{\rm \dagger}$.  
\begin{algorithm}[t]
	\begin{algorithmic} [1]
		\STATE Given : a lower triangular matrix  $\bGamma_{Q\times Q}$ and $\bX_{Q\times \theta}$
		\STATE Output : $\bY_{Q\times \theta}=\bGamma_{Q\times Q}^{-1}\bX_{Q\times \theta}$
		\FOR{$s=0:\theta$}
		\STATE $\bY(0,s)= \frac{\bX(0,s)}{\bGamma(0,0)}$
		\FOR{$k=1:\theta$}
		\STATE \small{$\bY(k,s)=\frac{1}{\bGamma(k,k)} {\bX(k,s)-\sum_{i=1}^{k-1}{\bGamma(k,k-i)\bY(k-i,s)}}$}
		\ENDFOR
		\FOR{$k=\theta +1:Q$}
		\STATE \small{$\bY(k,s)=\frac{1}{\bGamma(k,k)} {\bX(k,s)-\sum_{i=1}^{P-1}{\bGamma(k,k-i)\bY(k-i,s)}}$}
		\ENDFOR
		\ENDFOR
		\caption{Computation of $\bY=\bGamma^{-1}\bX $}
		\label{Alg:inverseoflowertriangularbandedmatrix}
	\end{algorithmic}
\end{algorithm}
As $\theta<< MN$, even a direct computation of (\ref{eqn:FG}) requires $O(\theta^2 MN)$ computations. As $\bF$ is a lower triangular matrix and $\bG$ is a upper triangular matrix, $\bF$ and $\bG$ can be computed using LU decomposition of (\ref{eqn:FG}). Pivotal Gaussian elimination algorithm \cite{golub2012matrix} can be used to compute LU decomposition of (\ref{eqn:FG})  without much increase in complexity. It should be noted that diagonal values of $\bbL$ and $\bF$ are unity. Thus, diagonal values of $\bL$ are also unity. 
	\subsubsection*{Note on the non-singularity of $\bL$ and $\bU$}
	For LMMSE processing, $\bL$ and $\bU$ need to be inverted (as will be discussed in Sec.~\ref{sec:lowcomplexity:computationofr}). We next discuss the non-singularity of $\bL$ and $\bU$. As $\HH$ is a hermitian matrix, its a positive semi-definite matrix. Since $\snri > 0$ for finite SNR ranges, $\channelequalizationmatrixop$ is a positive definite matrix; therefore, $\channelequalizationmatrixop$ is invertible.  As diagonal values of $\bL$ are unity, $\bL$ is  non-singular. Further, non-singularity of $\bU$  is a consequence of non-singularity of $\channelequalizationmatrixop$ \cite{golub2012matrix}. 
\subsection{Computation of $\br$} \label{sec:lowcomplexity:computationofr}
After $LU$ decomposition of $\channelequalizationmatrixop$, $\br_{ce}$ is simplified to,
\begin{equation}
\br_{ce}= \bH^{\rm \dagger} \overbrace{\bU^{-1} \underbrace{\bL^{-1} \br}}^{\br^{(2)}}_{\br^{(1)}}.
\end{equation}
As $\bL$ is a quasi-banded lower triangular matrix, $\br^{(1)}= \bL^{-1}\br$ can be computed using low complexity forward substitution as explained in Algorithm 2. $\br^{(2)}=\bU^{-1}\br^{(1)}$ can be computed using Algorithm 3.
\begin{algorithm}[t]
	\begin{algorithmic} [1]
		\STATE Given : a quasi banded lower triangular matrix  $\bL_{MN\times MN}$ and $\br_{MN\times 1}$
		\STATE Output : $\br^{(1)}_{MN\times 1}=\bL_{MN\times MN}^{-1}\br_{MN\times 1}$
		
		\STATE $\br^{(1)}(0)= \br(0)$
		\FOR{$k=1:\alpha-1$}
		\STATE $\br^{(1)}(k)={\br(k)-\sum_{i=1}^{k-1}{\bL(k,k-i)\br^{(1)}(k-i)}}$
		\ENDFOR
		\FOR{$k=\alpha:Q$}
		\STATE $\br^{(1)}(k)={\br(k)-\sum_{i=1}^{\alpha-1}{\bL(k,k-i)\br^{(1)}(k-i)}}$
		\ENDFOR
		\FOR{$k=Q+1:MN-1$}
		\STATE $\br^{(1)}(k)={\br(k)-\sum_{i=1}^{MN-1}{\bL(k,k-i)\br^{(1)}(k-i)}}$
		\ENDFOR
		\caption{Computation of $\br^{(1)}=\bL^{-1}\br$}
		\label{Alg:inverseoflowertriangularbandedmatrixquasibanded}
	\end{algorithmic}
	
\end{algorithm}

\begin{algorithm}[t]
	\begin{algorithmic} [1]
		\STATE Given : a quasi banded upper triangular matrix  $\bU_{MN\times MN}$ and $\br^{(1)}_{MN\times 1}$
		\STATE Output : $\br^{(2)}_{MN\times 1}=\bU_{MN\times MN}^{-1}\br^{(1)}_{MN\times 1}$
		
		\STATE $\br^{(2)}(MN-1)= \frac{\br^{(1)}(MN-1)}{\bU(MN-1,MN-1)}$
		\FOR{$k=MN-2:MN-2\alpha$}
		\STATE \small{$\br^{(2)}(k)=\frac{1}{\bU(k,k)} {\br^{(1)}(k)-\sum_{i=1}^{MN-k-1}{\bU(k,k+i)\br^{(2)}(k+i)}}$}
		\ENDFOR
		\FOR{$k=\alpha:Q$}
		\STATE $\br^{(2)}(k)=\frac{1}{\bU(k,k)} {\br^{(1)}(k)-\sum_{i=1}^{\alpha}{{\bU(k,k+i)\br^{(2)}(k+i)}}}$  $-\sum_{r=MN-\alpha}^{MN-1}{{\bU(k,r)\br^{(2)}(r)}} $
		\ENDFOR
		\caption{Computation of $\br^{(2)}=\bU^{-1}\br^{(1)}$}
		\label{Alg:inverseofuppertriangularbandedmatrix}
	\end{algorithmic}
	
\end{algorithm}
Using the definition of $\bH$, $\br_{ce}= \bH^{\rm \dagger} \br^{(2)}$ can be given as,
\begin{eqnarray}
\label{eqn:Hhermitian}
\br_{ce}&=& \sum_{p=1}^{P}{\bar{h}_p \dopplermatrix^{-\dopplerindex_p} \underbrace{\delaymatrix^{-\delayindex_p} \br^{(2)}}_{\text{circular shift}}} 
\end{eqnarray}
 To compute $\br_{ce}$, $\br^{(2)}$ is first circularly shifted by delay $-\delayindex_p$ and then multiplied by $\bar{h}_p \mathrm{diag}\{\dopplermatrix^{-\dopplerindex_p}\}$ by using point-to-point multiplication for each path $p$. All vectors obtained in above step are finally summed to obtain $\br_{ce}$. 
 
\begin{figure}[t]
	\centering
	\includegraphics[width=1\linewidth]{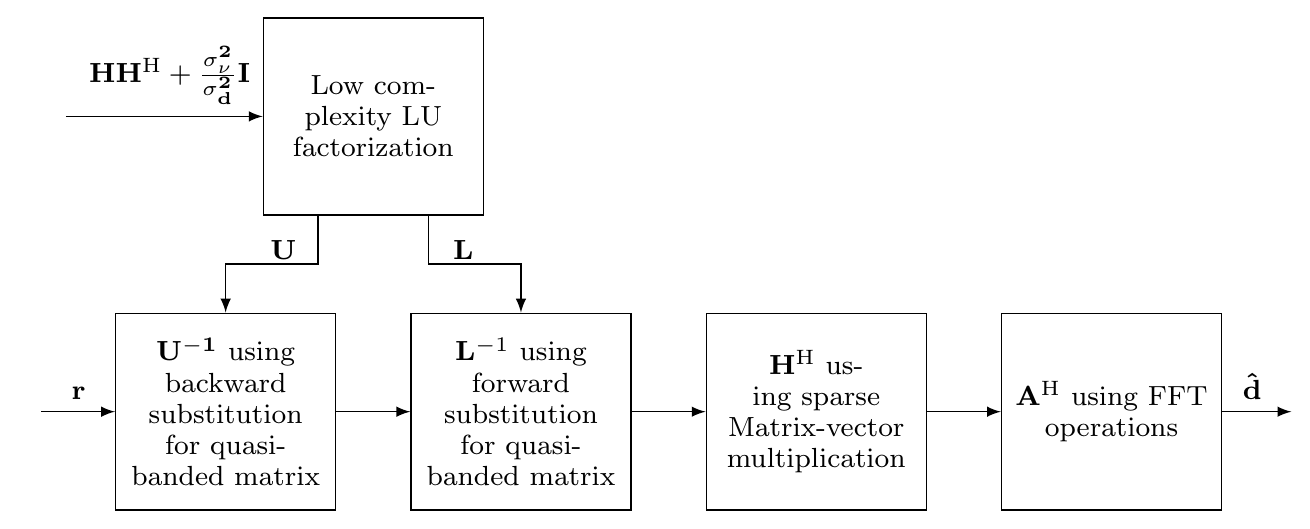}
	\caption{Our Proposed Low Complexity OTFS-MMSE Receiver}
	\label{fig:mmserxdiag}
\end{figure}

 Instead of directly computing $\hat{\bd}$ as $\bA^{\rm \dagger} \br_{ce}$, we first reshape $\br_{ce}$ to a $M\times N$ size $\bR$ matrix as,
 \begin{equation}
 \bR= \small{\begin{bmatrix}
 \br_{ce} (0) & \br_{ce}(M) & \cdots & \br_{ce}(MN-N) \\
 \br_{ce}(1) & \br_{ce}(M+1) & \cdots & \br_{ce}(MN-N+1) \\
 \vdots & \vdots & \ddots & \vdots  \\
 \br_{ce}(M-1) & \br_{ce}(2M-1) & \cdots & \br_{ce}(MN-1) \\
 \end{bmatrix}}.
 \end{equation}
 Then we perform
 \begin{equation}
 \label{eqn:oftsdemod}
 \hat{\bd}= vec\{\bR \bW_N^{\rm \dagger} \},
 \end{equation}
 which can be implemented using $M$ number of $N$-point FFT operations. Fig.~\ref{fig:mmserxdiag} describes the signal processing steps of our proposed low complexity LMMSE receiver.

\subsection{LMMSE receiver for OFDM over TVC } 
Low complexity receiver discussed for OTFS can easily be extended to OFDM by setting $\bA=\bI_N \kron \bW_M$. To do so, $\br_{ce}=\bH_{eq} \br$ is performed by computing (\ref{eqn:partitionofHH}-\ref{eqn:Hhermitian}) as discussed in Sec.~\ref{sec:lowcomplexity} and \ref{sec:lowcomplexity:computationofr}. Further, $\hat{\bd}=(\bI_N \kron \bW_M^{\dagger}) \br_{ce}$ can be computed using $N$ number of $M$-point FFTs. 
\section{Result}
\subsection{Computational Complexity}
In this section, we  present the computational complexity of our proposed LMMSE receiver. We calculate the complexity in terms of total number of complex multiplications (CMs). $c$-point FFT and IFFT can be implemented using radix-2 FFT algorithm using $\frac{c}{2}\log_2(c)$ CMs \cite{blahut_fast_2010}. The complexity of the proposed receiver can be computed using the structure provided in Sec.~\ref{sec:lowcomplexity}. Computation of $c\times c$ matrix-matrix multiplication, matrix inversion and LU decomposition require $\frac{c^3}{2}$, $\frac{2c^3}{3}$ and $\frac{2c^3}{3}$ CMs respectively. 
\begin{table}[t]
	\centering 
		\caption{Computational complexity of different operations in our proposed receiver}
		\label{tab:complexitydifferentoperations}
	\begin{tabular}{|m{0.45\linewidth}|m{0.45\linewidth}|}
		\hline
	\textbf{Operation} & \textbf{Number of Complex Multiplications}\\ \hline 
	(\ref{eqn:closedformchest}) & $[P^2-P][2\beta+1]MN+P$ \\ \hline
	(\ref{eqn:T}) & $[\alpha^2+2\alpha]MN$ \\ \hline
	(\ref{eqn:E},\ref{eqn:V}) using Algorithm ~\ref{Alg:inverseoflowertriangularbandedmatrix} & $\alpha MN-\frac{3\alpha^3+\alpha}{2}$\\ \hline
	(\ref{eqn:FG}) and LU decomposition of $\bF \bG$ & $\alpha^2 MN-MN+\frac{2\alpha^3}{3}$ \\ \hline
	Algorithm 2 and 3 & $MN[2\alpha -1]+\frac{3\alpha^2}{2}+\frac{\alpha}{2}$ \\ \hline
	(\ref{eqn:Hhermitian}) & $P(\beta+1)MN$\\ \hline
	(\ref{eqn:oftsdemod}) & $\frac{MN}{2} \log_2(N)$ \\ \hline
	\end{tabular}
\end{table}
Total CMs required to compute different operations in our receiver are presented in  Table~\ref{tab:complexitydifferentoperations}. CMs required for different receivers is presented in Table~\ref{tab:complexity}. It is evident that the our proposed receiver has complexity of $O(MN[\log_2(N)+\alpha^2+P^2 \beta])$. 
\begin{table}[t]
	\centering
	\caption{Computational complexity of different receivers}.
	\label{tab:complexity}
	\begin{tabular}{|m{0.3\linewidth}|m{0.6\linewidth}|}
		\hline
		\textbf{Structure} & \textbf{Number of Complex Multiplications}\\ \hline 
		OFDM receiver direct using (\ref{eqn:MMSEOTFS})& $\frac{MN}{2}\log_2M+ \frac{8}{6}(MN)^3+2(MN)^2$ \\ \hline \vspace{0.1cm}
	    OTFS receiver direct using (\ref{eqn:MMSEOTFS}) & $\frac{MN}{2}\log_2N+ \frac{8}{6}(MN)^3+2(MN)^2$ \\ \hline \vspace{0.1cm}
		our proposed OFDM receiver & $\frac{MN}{2}\log_2M+ MN [2\alpha^2+2P^2\beta+9\alpha-P\beta -3]+\frac{2}{3}\alpha^3+2\alpha+P$ \\ \hline \vspace{0.1cm}    
			our proposed OTFS receiver & $\frac{MN}{2}\log_2N+  MN [2\alpha^2+2P^2\beta+9\alpha-P\beta -3]+\frac{2}{3}\alpha^3+2\alpha+P$ \\ \hline  
	\end{tabular}
\end{table}

\begin{figure}[h]
	\centering
	\includegraphics[width=0.9\linewidth]{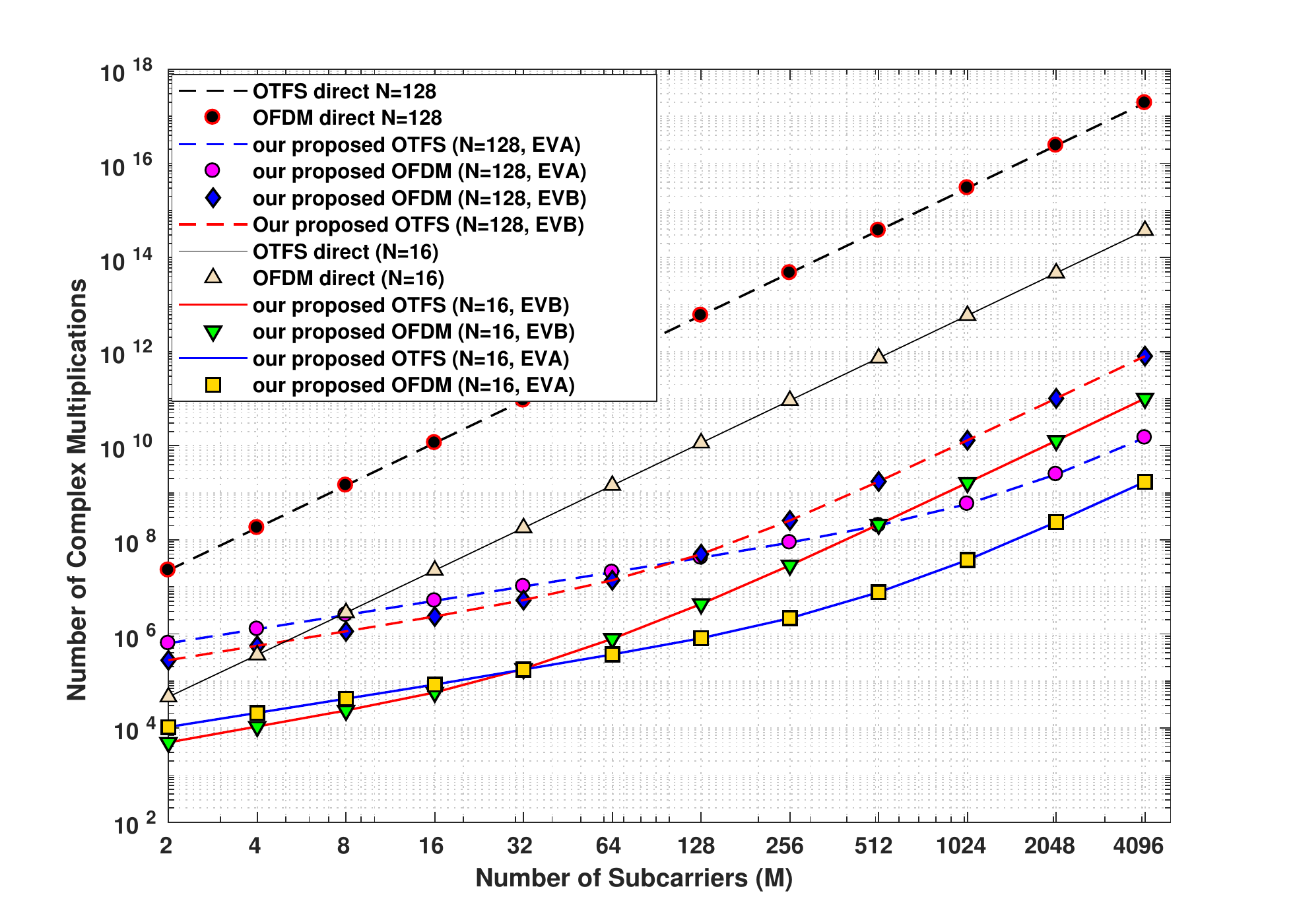}
	\caption{Computation complexity comparison of different receivers.  }
	\label{fig:complexityeva}
\end{figure}
To evaluate the complexity reduction achieved by our proposed receiver, we consider an OTFS system with $\subcarrier=15~ KHz$, $f_c=4~GHz$ and vehicular speed of 500 kmph. We consider two 3GPP vehicular channel models \cite{series2009guidelines} namely (i) extended vehicular A (EVA) with $P=9$ and $\tau_{max}=2.51 ~\mu~sec$ , and (ii) extended vehicular B (EVB) with $P=6$ and $\tau_{max}=20 ~\mu~sec$. Two block durations are assumed namely (i) small block with $N=16$ and $T_f=1.1 ~msec.$ and (ii) large block with $N=128$ and $T_f=8.85~msec$. 
The complexity presented in Table~\ref{tab:complexity} is plotted in Figure~\ref{fig:complexityeva} for  $M\in[2 ~ 4096]$. It is evident from the figure that for EVA channel our proposed receivers require up-to $10^7$ and $10^5$ times lower CMs than direct ones using (\ref{eqn:MMSEOTFS}) for large and small block respectively. 
	Whereas for EVB channel, our proposed receiver need $2.5\times 10^5$ and $3000$ times lesser CMs over the direct ones using  (\ref{eqn:MMSEOTFS}) for large and small block respectively. This reduction in complexity gain for EVB channel as compared with EVA channel is due to increase in $\alpha$. We can conclude that our proposed receivers achieve a significant complexity reduction over direct implementation of (\ref{eqn:MMSEOTFS}).

\subsection{BER Evaluation}
  
\begin{table} [h]
	\centering
	\caption{Simulation Parameters}
	\label{tab:simu:para:mmse}
	\begin {tabular}{|m{0.45\linewidth}|m{0.45\linewidth}|}
	\hline
	Number of Sub-carriers $M$ & 512\\ \hline
	Number of Time-slots $N$ & 128 \\ \hline
	Mapping & 4 QAM \\ \hline
	Sub-carrier Bandwidth & 15 KHz \\ \hline
	Channel & EVA \cite{series2009guidelines} \\ \hline
	Vehicular Speed& 500 Kmph   \\ \hline
	Carrier Frequency & 4 GHz \\ \hline
\end{tabular}
\end{table}
\begin{figure}[h]
	\centering
	\includegraphics[width=\linewidth]{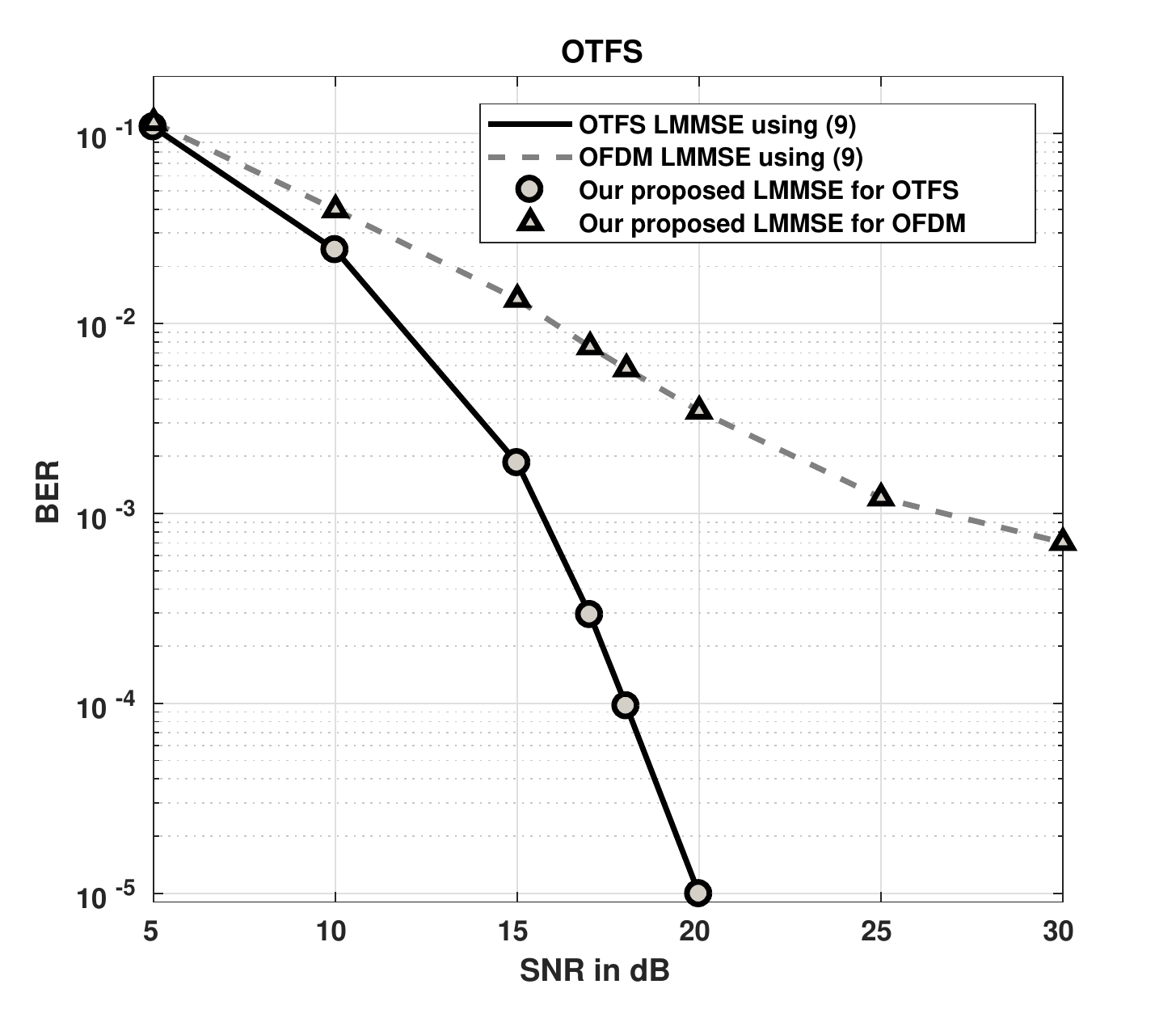}
	\caption{BER comparison of our proposed receiver with the direct one using (\ref{eqn:MMSEOTFS}) for 4 QAM modulation.}
	\label{fig:patentber}
\end{figure}
Here we present BER performance of the proposed receiver in EVA channel. Simulation parameters are given in Table~\ref{tab:simu:para:mmse}. Doppler is generated using Jake's formula, $\nu_p=\nu_{max} cos(\theta_p)$, where $\theta_p$ is uniformly distributed over $[-\pi ~ \pi]$. The CP is chosen long enough to accommodate the wireless channel delay spread.
 Figure~\ref{fig:patentber} compares BER performance of our proposed receiver with the direct ones using (\ref{eqn:MMSEOTFS}). It can be observed that the proposed receiver does not suffer from any performance degradation when compared with the direct ones. It can also be observed that OTFS-LMMSE receiver can extract diversity gain, for instance at the BER of $5\times 10^{-4}$, OTFS-LMMSE receiver achieves an SNR gain of 13 dB over OFDM-MMSE receiver.

\section{Conclusion}
In this paper, we have proposed a low complexity LMMSE receiver for OTFS waveform. The proposed technique exploit sparsity and quasi banded structure of matrices involved in LMMSE processing without incurring any performance penalty. We have shown that our proposed receiver can achieve upto $10^7$ times complexity reduction over direct implementation. Such substantial reduction with linear receiver is expected to provide an impetus for practical realization of future wireless OTFS based systems.
\bibliographystyle{IEEEtran}
\bibliography{OTFS,REF,manual,Complexity,OFDM,manual}
\end{document}